\documentclass[preprint,aps]{revtex4}

\usepackage{graphicx}

\begin{document}

\title{Weak Electron-Phonon Coupling and Unusual Electron Scattering of Topological Surface States in Sb(111) by Laser-Based Angle-Resolved Photoemission Spectroscopy}
\author{Zhuojin Xie$^{1}$, Shaolong He$^{1,*}$, Chaoyu Chen$^{1}$, Ya Feng$^{1}$, Hemian Yi$^{1}$, Aiji Liang$^{1}$, Lin Zhao$^{1}$, Daixiang Mou$^{1}$, Junfeng He$^{1}$, Yingying Peng$^{1}$, Xu Liu$^{1}$, Yan Liu$^{1}$, Guodong Liu$^{1}$, Xiaoli Dong$^{1}$, Li Yu$^{1}$, Jun Zhang$^{1}$, Shenjin Zhang$^{3}$, Zhimin Wang$^{3}$, Fengfeng Zhang$^{3}$, Feng Yang$^{3}$, Qinjun Peng$^{3}$, Xiaoyang Wang$^{3}$, Chuangtian Chen$^{3}$, Zuyan Xu$^{3}$ and X. J. Zhou$^{1,2,*}$}

\affiliation{
\\$^{1}$Beijing National Laboratory for Condensed Matter Physics, Institute of Physics, Chinese Academy of Sciences, Beijing 100190, China
\\$^{2}$Collaborative Innovation Center of Quantum Matter, Beijing, China
\\$^{3}$Technical Institute of Physics and Chemistry, Chinese Academy of Sciences, Beijing 100190, China
}
\date{May 4, 2014}
\begin{abstract}
High resolution laser-based angle-resolved photoemission (ARPES) measurements have been carried out on Sb(111) single crystal. Two kinds of Fermi surface sheets are observed that are derived from the topological surface states: one small hexagonal electron-like Fermi pocket around $\Gamma$ point and the other six elongated lobes of hole-like Fermi pockets around the electron pocket.  Clear Rashba-type band splitting due to the strong spin-orbit coupling is observed that is anisotropic in the momentum space.  Our super-high-resolution ARPES measurements reveal no obvious kink in the surface band dispersions indicating a weak electron-phonon interaction in the surface states. In particular, the electron scattering rate for these topological surface states is nearly a constant over a large energy window near the Fermi level that is unusual in terms of the conventional picture.
\end{abstract}


\maketitle

Topological insulators have attracted much attention recently because of their unique electronic structure, spin texture and potential applications ranging from spintronic devices to topological quantum computation~\cite{Hasan2010,Qi2011}. Topological insulators have an insulating energy gap in the bulk, and gapless edge or surface states that are protected by the time-reversal symmetry~\cite{Hasan2010,Qi2011}. The robust metallic surface states in three-dimensional topological insulators have an odd number of Fermi level crossings; this is a  main characteristic of the nontrivial topological order. In addition, electron spins in the surface states are locked with the crystal momentum forming chiral spin states~\cite{Hsieh2009a}, and can also be locked with the atomic orbitals~\cite{Zhang2013,xie2014}.  Antimony (Sb) is a semimetal that has an indirect negative gap between the valence band maximum at the H-point of the bulk Brillouin zone (BZ) and the conduction band minima at three equivalent L-points~\cite{golin,liu1995}.  It has been found experimentally that Sb(111) surface states have an odd number of bands crossings the Fermi level along the high-symmetry directions that are highly spin-polarized~\cite{Sugawara2006,Hsieh2009}. Therefore, the surface states of Sb(111) are expected to be topologically nontrivial although the bulk is a semimetal. Study of the pure Sb will be helpful in understanding the topological order in the Bi$_{1-x}$Sb$_x$ series~\cite{Hsieh2009}. It has been suggested that the quantum spin Hall effect can be achieved in the spin-orbit-coupling materials without external magnetic fields, with Bi(111) and Sb(111) as good candidates~\cite{Bernevig2006,Konig2007}. The realization of novel quantum phenomena depends sensitively on the electron scattering of the surface states so the study of the related many-body effects becomes necessary and important\cite{CYChenSR}.

In this Letter, we investigate the electronic structure and many-body effects of Sb(111) crystals by using our vacuum ultraviolet (VUV) laser-based angle-resolved photoemission (ARPES) system. We have clearly observed two kinds of Fermi surface sheets derived from the surface states.  One is a small hexagonal electron-like Fermi pocket at the $\Gamma$ point and the other is six elongated hole-like Fermi pockets around the electron pocket. Clear Rashba-type splitting of the surface bands has been observed which is anisotropic in the momentum space.   The systematic study on the many-body effects of these Rashba-split bands indicate that the electron-phonon coupling in the surface states of Sb(111) is quite weak. The imaginary part of the electron self-energy of the surface bands is nearly constant over a wide energy range ($\sim$100 meV) near the Fermi level, indicating an unusual electron scattering of the Sb(111) surface states.

We have grown high-quality Sb single crystals by using the flux method and the traveling solvent floating zone method. High-quality Sb single crystal with a length of 5 cm is obtained (Fig. 1, upper-left inset).  Small  plate-like samples with a typical size of $~2\times3\times0.1 mm^3$ can be obtained from these large pieces. The crystal structure and the quality of the obtained Sb crystals were characterized by the X-ray diffraction (XRD) (Fig. 1).  Sb has an A7 crystal structure with a rhombohedral unit cell that contains two Sb atoms (upper-right inset in Fig. 1)\cite{liu1995}.  The rhombohedral structure can be thought of as a simple cubic structure stretched along its diagonal axis; this axis then becomes the trigonal (111) axis and retains three-fold symmetry~\cite{yang1999,Wyckoff1963}.  This trigonal axis of the Sb crystal structure corresponds to the (003) direction in the hexagonal crystallographic structure~\cite{Wyckoff1963,Jezequel1997} and the (111) direction in the rhombohedral crystallographic structure~\cite{yang1999,Wyckoff1963}. As seen in Fig. 1, all the diffraction peaks of the cleaved Sb single crystal can be indexed into $(0,0,h)$ in the hexagonal crystallographic structure, with $h$ being 3, 6 and 9. The c-axis lattice constant is determined to be $c=11.223$ \AA  which is consistent with the result reported before~\cite{golin,liu1995}.  The sample cleavage plane is then (003)Hex/(111)Rhom, denoted as Sb(111) hereafter.

The angle-resolved photoemission measurements were performed on our newly developed VUV laser-based spin- and angle-resolved photoemission system (SARPES)~\cite{xie2014,gdliu}, which combines the Scienta R4000 analyzer with  Mott-type spin detectors. The photon energy of the laser is 6.994 eV with a bandwidth of 0.26 meV. The best energy resolution for regular ARPES measurements is $\sim$ 1 meV. The angular resolution is $\sim$ 0.3$^{\circ}$, corresponding to a momentum resolution 0.004\AA$^{-1}$ for the 6.994 eV photon energy. The single crystal Sb samples were cleaved and {\it in situ} measured in an ultra-high vacuum with a base pressure better than 5$\times$10$^{-11}$ Torr.

Figure 2 shows the Fermi surface mapping of Sb(111). In the mapping covering a large momentum space (Fig. 2a), there is a small Fermi pocket around $\Gamma$ point, and six lobes of Fermi pockets surrounding the central Fermi pocket. Near M, there is the third oval-shaped Fermi pocket. The band structure measurements along high symmetry cuts (Figs. 2c and 2d) indicate that the central Fermi pocket is electron-like while the six lobes are hole-like. Fig. 2b shows a high resolution measurements of the central electron-like Fermi pocket that shows a clear hexagon-like shape. It is noted that, while the electron-like Fermi pocket near $\Gamma$ shows nearly a six-fold symmetry, the spectral weight distribution of the six hole-like lobes has an obvious three-fold symmetry (Fig. 2a). It is known that the bulk Sb has three-fold symmetry and the Sb(111) surface has six-fold symmetry. Therefore, the central electron-like Fermi pocket should have predominantly surface origin. But the six hole-like Fermi pockets must have a strong mixing between the surface state and the bulk state. Our laser-ARPES results (6.994 eV photon) are consistent with the previous results from the helium lamp and synchrotron radiation~\cite{Sugawara2006,Hsieh2009}, reaffirming the nature of the low-binding-energy bands as mainly from the surface states.

Our laser-ARPES measurements clearly show Rashba-type band splitting due to strong spin-orbit coupling. Figures 2c and 2d show band structures measured along high symmetry cuts of Sb(111). For the $\Gamma$-M cut (Fig. 2c), there are four bands (labeled as M1, M2£¬ M3 and M4 in Fig. 2c) observed that all cross the Fermi level. However, for the $\Gamma$-K cut (Fig. 2d), there are only two bands (labeled as K2 and K3 in Fig. 2d) that cross the Fermi level while the other two bands (K1 and K4) bend back at a binding energy of 0.1 eV and a momentum of $\sim$$\pm$$0.1($\AA$^{-1})$) and penetrate into the bulk states. In this case, the bands between the Fermi level and 0.2 eV binding energy are mainly from surface states while the bands above 0.2 eV binding energy are of bulk nature. The band structure along $\Gamma$-M is quite similar to that of Au(111) surface state where, because of Rashba effect, four spin-polarized bands are formed\cite{LaShell1996,Reinert2001,MHoesch}.  The four bands of Sb(111) along $\Gamma$-M are also spin-polarized\cite{Hsieh2009} and they are formed because of the strong spin-orbit coupling, the same origin as that in Au(111). Interestingly, the Rashba splitting in Sb(111) along $\Gamma$-M is stronger than that in Au(111) although Au (Z=79) is heavier than Sb (Z=51).

The quite different behaviors between $\Gamma$-M and $\Gamma$-K cuts indicates a strong anisotropy of Rashba splitting in Sb(111). To keep track on the band evolution, Fig. 3 shows the band structures of Sb(111) along many momentum cuts in between the $\Gamma$-M and $\Gamma$-K directions. In the high resolution Fermi surface mapping (Fig. 3a), we clearly resolve the central hexagonal electron-like Fermi pocket, and a complete lobe of the hole-like Fermi pockets.  The inner bands (M2, M3) along $\Gamma-M$ direction transform naturally into K2 and K3 bands along the $\Gamma$-K direction.  These inner bands form the hexagonal electron-like Fermi pocket centered at $\Gamma$ point. On the other hand, along $\Gamma$-K,  the  outer band (K4) bend back towards the bulk valence band near a momentum of $\sim$0.1 \AA$^{-1}$ for the $\Gamma$-K cut (cut 1). With the momentum cuts rotating away from $\Gamma$-K, the bending point and mixing point of the outer band gradually goes away from the $\Gamma$ point with its bending energy getting closer to the Fermi level (Fig. 3(b1-b7)). For the cut 6, one can see three Fermi crossings, one on the electron-like Fermi pocket, and two on the lobe of a hole-like Fermi pocket. It is clear that the outer bands participate in generating the lobe of hole-like Fermi pockets.

Now we come to analyze the electron dynamics of the surface states that can be realized by extracting the electron self-energy\cite{XJZhouReview}. The utilization of laser-ARPES is particularly advantageous in this regard because of the super-high energy resolution, high momentum resolution and high data statistics\cite{gdliu,CYChenSR}. Fig. 4 shows the results for the surface state bands that cross the Fermi level along $\Gamma$-M and $\Gamma$-K high symmetry directions. Fig. 4c and 4d show the momentum distribution curves (MDCs) corresponding to the measured band structures (Fig. 4a and 4b) along $\Gamma$-M and $\Gamma$-K, respectively. The MDCs are fitted to get the band dispersions (Fig. 4e and 4f) and the MDC width from which the real part of the electron self-energy (Fig. 4g) and imaginary part of the electron self-energy (Fig. 4h) can be obtained\cite{XJZhouReview}.  We note that the obtained value of the imaginary part of electron self-energy depends on the instrumental resolution and sample inhomogeneity. In some optimal conditions, we have observed even smaller imaginary parts of the electron self-energy.

The measured band dispersions for the six surface bands (Fig. 4e and 4f) are nearly linear. We do not see obvious kink in these dispersions. In the corresponding effective real part of the electron self-energy (Fig. 4g), within the noise level of $\pm$(1$\sim$2) meV, the real part of the electron self-energy is basically zero within the [E$_F$, 0.1eV] energy range for all the six surface bands. No clear peak is observed from these real parts of the electron self-energy. These results indicate that the electron coupling with other collective excitations like phonons is extremely weak. This is similar to the case of the p-type Bi$_2$Te$_3$\cite{CYChenSR}. Our results differ from the previous results where a kink at $\sim$12 meV was reported\cite{Sugawara2006} although our instrumental resolution and data quality are much improved, and also the imaginary part of the electron self-energy is smaller than that in \cite{Sugawara2006}.

The imaginary part of the electron self-energy for the six surface bands (Fig. 4h) show some peculiar behaviors. First, we find that different surface bands may exhibit different imaginary part of election self-energy, i.e., different bands may experience different electron scatterings. Second, the energy-dependence of the scattering rate is quite unusual for the Sb(111) surface states. Except for the M4 band where the imaginary part of the electron self-energy shows a slight decrease with decreasing binding energy, it is nearly a constant over a large energy window of 100 meV for M3, K1 and K2 bands. For M1 and M2 bands, the imaginary part of the electron self-energy even slightly increases with decreasing binding energy. The contributions of the imaginary part of electron self-energy usually come from electron-electron interaction, electron-phonon coupling and electron-impurity (disorder) scattering\cite{XJZhouReview}. The imaginary part of the electron self-energy is expected to increase with increasing binding energy, due to the electron-electron interaction, as has been experimentally observed in simple metals\cite{TVallaMo}, in the topological insulators\cite{CYChenSR}, and in the high temperature cuprate superconductors\cite{XJZhouPRL}. In this sense, the electron scattering rate of the surface bands in Sb(111) is quite unusual. Whether strong spin-orbit coupling and/or mixing of the bulk and surface states can account for such a behavior needs further investigations.

In summary, we have investigated the Fermi surface, band structure and many-body effects of Sb(111) single crystal by using our VUV laser-based spin- and angle-resolved photoemission system. Rashba-type band splitting is observed and it is anisotropic in the momentum space. We do not observe obvious kink in the dispersions of the surface state bands indicating a rather weak electron-phonon coupling. In particular, we have observed unusual electron scattering behaviors in the surface states of Sb(111). The present work also demonstrates the feasibility of laser ARPES in studying Sb, opening a window for further study of the electronic structure and spin texture in the Bi$_{1-x}$Sb$_{x}$ series by using laser-based spin- and angle-resolved photoemission system\cite{xie2014}.

This work is supported by the National Natural Science Foundation of China (91021006 and 11174346) and the Ministry of Science and Technology of China (2011CB921703, 2013CB921700 and 2013CB921904).

$^{*}$Corresponding author: XJZhou@aphy.iphy.ac.cn, shaolonghe@aphy.iphy.ac.cn

\newpage

\begin{figure*}[tbp]
\begin{center}
\includegraphics[width=1.0\columnwidth,angle=0]{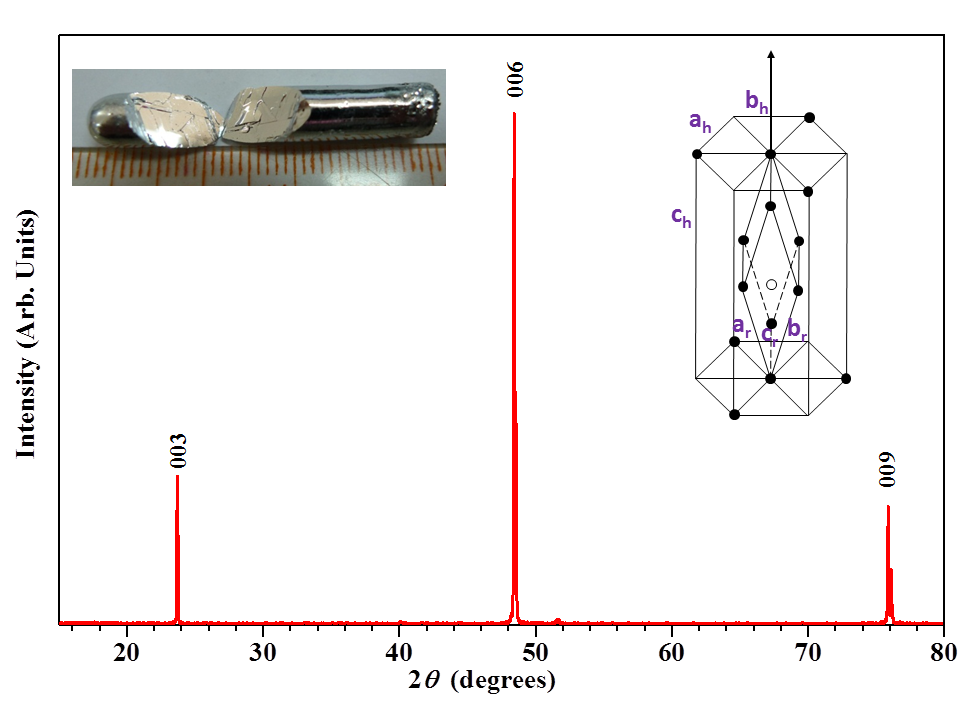}
\end{center}
\caption{X-ray diffraction analysis of the Sb single crystals. The XRD pattern was obtained by using a rotating anode Cu $K_{\alpha}$ radiation ( $\lambda$= 1.5418 \AA) under $\theta \sim 2\theta$ scan mode. The upper-left inset shows a photograph of as-grown Sb ingots from which small single crystals can be obtained that are plate-like with a typical size of 2$\sim$3 mm with a thickness of $\sim$0.1 mm. The upper-right inset shows the crystal structure of antimony where $\bf{a_r}$, $\bf{b_r}$ and $\bf{c_r}$ are the abc planes of the rhombohedral structure and $\bf{a_h}$, $\bf{b_h}$ and $\bf{c_h}$ are the abc planes of the hexagonal structure.  All the measured peaks can be indexed into $(0,0,h)$ in the hexagonal crystallographic structure.
}
\end{figure*}

\begin{figure*}[tbb]
\begin{center}
\includegraphics[width=1.0\columnwidth,angle=0]{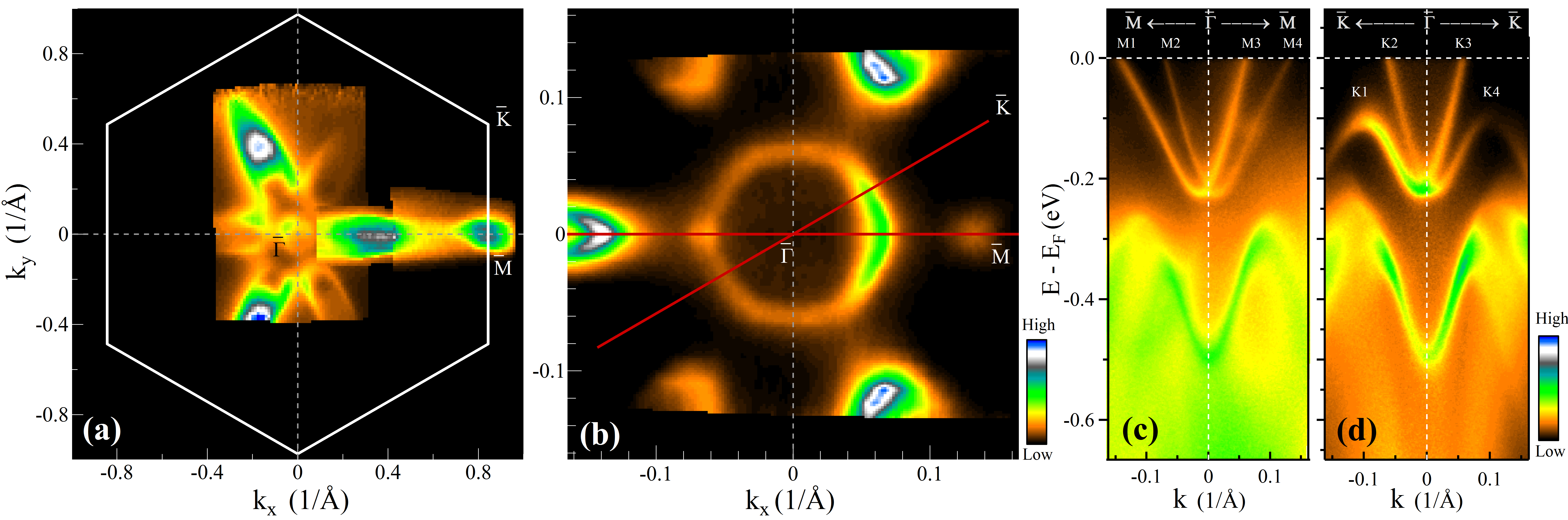}
\end{center}
\caption{ Fermi Surface of Sb(111) and the band structures along high-symmetry $\Gamma-M$ and $\Gamma-K$ directions. (a). Fermi surface mapping of Sb(111) over a large momentum space. In addition to a small Fermi pocket near $\Gamma$, there are six lobes of Fermi pockets around $\Gamma$ point. (b). Zoom-in Fermi surface mapping near the $\Gamma$ point. A small hexagonal Fermi pocket can be clearly seen.  (c). Band structure of Sb(111) measured along $M-\Gamma-M$ cut (shown in (b) as red line).  (d). Band structure of Sb(111) measured along $K-\Gamma-K$ cut (shown in (b) as red line).
}
\end{figure*}

\begin{figure*}[tbp]
\begin{center}
\includegraphics[width=1.0\columnwidth,angle=0]{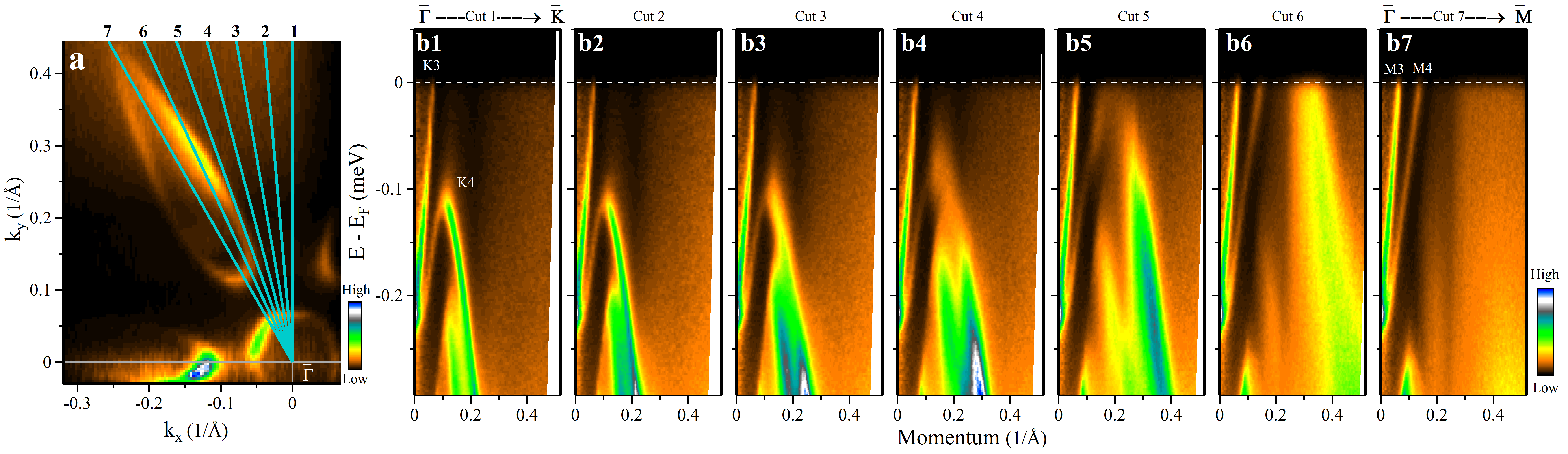}
\end{center}
\caption{Band structure evolution between high symmetry directions $\Gamma-K$ and $\Gamma-M$. (a). Fermi surface mapping covering both the central hexagonal electron pocket and a complete lobe of the hole pockets. The location of the seven momentum cuts (blue solid lines) are marked by numbers from 1 to 7.   (b). Band structure evolution between high symmetry directions $\Gamma-K$ (cut 1) and $\Gamma-M$ (cut 7).}
\end{figure*}

\begin{figure*}[tbp]
\begin{center}
\includegraphics[width=1.0\columnwidth,angle=0]{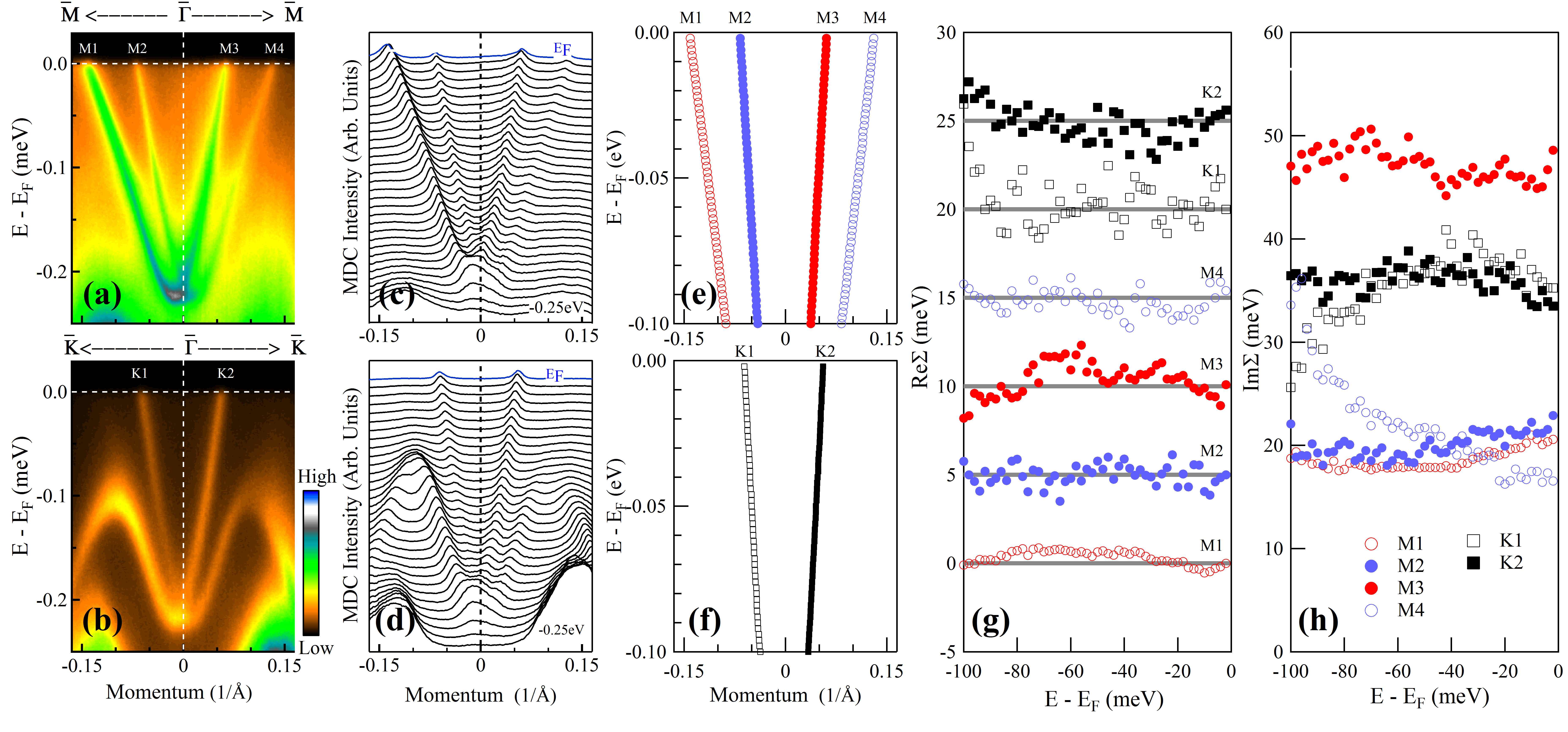}
\end{center}
\caption{ Dispersion and effective electron self-energy of the Sb(111) surface state bands along $\Gamma-K$ and $\Gamma-M$ directions. (a) and (b) show band structures measured along $\Gamma-M$ and $\Gamma-K$ directions, respectively. (c) and (d) show the corresponding momentum distribution curves (MDCs) for the two images. Blue lines indicate the MDCs at the Fermi level ($E_F$) and the bottom MDCs correspond to a binding energy of 0.25 eV.  (e) and (f) show the band dispersions in a small energy window of [0,-100meV] obtained from the MDC fitting.  (g). The effective real part of the electron self-energy $Re\Sigma(k, \omega)$ for the 6 bands shown in (a) and (b). For clarity, the curves are offset by 5, 10, 15, 20 and 25 meV for M2, M3, M4, K1 and K2 bands, respectively. The bare bands are chosen for each band as straight lines connecting the two points on the measured dispersion at the Fermi level and 0.1 eV binding energy.  (h). The imaginary part of the electron self-energy $Im\Sigma(k, \omega)$ for the 6 bands shown in (a) and (b).
}
\end{figure*}

\end{document}